\documentclass[12pt]{jpconf}
\usepackage{iopams}
\usepackage{bm}
\usepackage{graphicx}
\usepackage{cite}
\usepackage{multirow}
\usepackage{mathtools}
\usepackage{url}
\usepackage[hyperindex]{hyperref}
\usepackage{subfigure}
\usepackage{stackrel}
\usepackage{paralist}
\usepackage{xspace}
\newcommand{\nua}[1]{\ensuremath{\rlap{\kern-2.5pt\ensuremath{\overset{\scriptscriptstyle(-)}{\phantom{\nu}}}}{\ensuremath{{\nu}_{#1}}}}\xspace}

\begin{document}

\title{Oscillations Beyond Three-Neutrino Mixing}

\author{Carlo Giunti}

\address{INFN, Sezione di Torino, Via P. Giuria 1, I--10125 Torino, Italy}

\ead{giunti@to.infn.it}

\begin{abstract}
The current status of the phenomenology
of short-baseline neutrino oscillations
induced by light sterile neutrinos in the framework of 3+1 mixing is reviewed.
\end{abstract}

The current experimental and theoretical research
of the physics of massive neutrinos is
based on the standard paradigm of three-neutrino ($3\nu$) mixing
which describes the oscillations of neutrino flavors
measured in solar, atmospheric and long-baseline experiments
\cite{Forero:2014bxa,Gonzalez-Garcia:2014bfa,Capozzi:2016rtj}.
In this framework,
the three left-handed active neutrino fields
$\nu_{eL}$,
$\nu_{\mu L}$,
$\nu_{\tau L}$
are unitary linear combinations of three left-handed massive neutrino fields
$\nu_{1L}$,
$\nu_{2L}$,
$\nu_{3L}$
with respective masses
$m_1$,
$m_2$,
$m_3$:
\begin{equation}
\nu_{\alpha L}
=
\sum_{k=1}^{N}
U_{\alpha k} \nu_{kL}
\qquad
(\alpha=e,\mu,\tau)
,
\label{mixing}
\end{equation}
for $N=3$,
where $U$ is the $3\times3$ unitary mixing matrix.
There are two independent squared-mass differences:
the small solar squared-mass difference
\begin{equation}
\Delta{m}^2_{\text{SOL}}
=
\Delta{m}^2_{21}
\simeq
7.4 \times 10^{-5} \, \text{eV}^2
,
\label{SOL}
\end{equation}
and the larger atmospheric squared-mass difference
\begin{equation}
\Delta{m}^2_{\text{ATM}}
=
|\Delta{m}^2_{31}|
\simeq
|\Delta{m}^2_{32}|
\simeq
2.5 \times 10^{-3} \, \text{eV}^2
,
\label{ATM}
\end{equation}
with
$\Delta{m}^2_{kj}=m_k^2-m_j^2$.

Although the standard $3\nu$ framework is very successful at explaining the
currently well-established neutrino data,
it is interesting to explore non-standard effects in neutrino oscillations,
which are expected from the new physics beyond the Standard Model.
In this review we consider the current short-baseline neutrino oscillation anomalies
and we discuss their explanation in the framework of the 3+1 mixing scheme.
There are three short-baseline neutrino oscillation anomalies:
\begin{enumerate}

\renewcommand{\labelenumi}{\theenumi.}
\renewcommand{\theenumi}{\arabic{enumi}}

\item
The
LSND observation of a short-baseline excess of
$\bar\nu_{e}$-induced events
in a $\bar\nu_{\mu}$ beam
\cite{Athanassopoulos:1995iw,Aguilar:2001ty}.

\item
The Gallium neutrino anomaly
\cite{Abdurashitov:2005tb,Laveder:2007zz,Giunti:2006bj,Giunti:2010zu,Giunti:2012tn},
consisting in a short-baseline disappearance of $\nu_{e}$
measured in the
Gallium radioactive source experiments
GALLEX
\cite{Kaether:2010ag}
and
SAGE
\cite{Abdurashitov:2009tn}.

\item
The reactor antineutrino anomaly
\cite{Mention:2011rk},
which is a deficit of the rate of $\bar\nu_{e}$ observed in several
short-baseline reactor neutrino experiments
in comparison with that expected from the calculation of
the reactor neutrino fluxes
\cite{Mueller:2011nm,Huber:2011wv}.

\end{enumerate}

A neutrino oscillation explanation of these anomalies
requires the existence of at least one additional squared-mass difference
\begin{equation}
\Delta{m}^2_{\text{SBL}} \gtrsim 1 \, \text{eV}^2
,
\label{SBL}
\end{equation}
which is much larger than
$\Delta{m}^2_{\text{ATM}}$
and
requires the existence of at least one massive neutrino
$\nu_4$
in addition to the three standard massive neutrinos
$\nu_1$,
$\nu_2$,
$\nu_3$
(see the review in Ref.~\cite{Gariazzo:2015rra}).
Since from the LEP measurement of the invisible width of the $Z$ boson
we know that there are only three active neutrinos,
in the flavor basis the additional massive neutrinos correspond to
sterile neutrinos
\cite{Pontecorvo:1968fh},
which do not have standard weak interactions.

In the general case of $N>3$ massive neutrinos,
the mixing of the three active neutrino fields
which are observable through weak interactions
is given by Eq.~(\ref{mixing})
with $N\geq4$ and $U$ is a $3 \times N$ rectangular mixing matrix
which is obtained by keeping only the first three rows of a
unitary $N \times N$ unitary matrix.
Moreover,
the mixing of the standard active neutrinos with the non-standard massive neutrinos
must be very small,
in order not to spoil the successful $3\nu$ mixing explanation of
solar, atmospheric and long-baseline
neutrino oscillation measurements
\cite{Forero:2014bxa,Gonzalez-Garcia:2014bfa,Capozzi:2016rtj}:
\begin{equation}
|U_{\alpha k}|^2 \ll 1
\qquad
(\alpha=e,\mu,\tau; \, k=4, \ldots, N)
.
\label{smallmix}
\end{equation}
In other words,
the non-standard massive neutrinos must be mostly sterile.

In this review we consider the so-called 3+1 scheme\footnote{
The 2+2 mixing schemes which were favorite in the 90's after the results of the LSND experiment
\cite{Okada:1996kw,Bilenky:1996rw,Bilenky:1999ny,Grimus:2001mn}
are excluded by solar and atmospheric neutrino oscillation data
\cite{Maltoni:2002xd,Maltoni:2004ei}.
In the literature one can also find studies of the
3+2
\cite{Sorel:2003hf,Karagiorgi:2006jf,Maltoni:2007zf,Karagiorgi:2009nb,Blennow:2011vn,Giunti:2011gz,Donini:2012tt,Archidiacono:2012ri,Conrad:2012qt},
3+3
\cite{Maltoni:2007zf,Rahman:2015hna},
3+1+1
\cite{Nelson:2010hz,Fan:2012ca,Kuflik:2012sw,Huang:2013zga,Giunti:2013aea},
and
1+3+1
\cite{Kopp:2011qd,Kopp:2013vaa}
schemes.
}
in which there is a non-standard massive neutrino (mostly sterile) at the eV scale
which generates the new squared-mass difference in Eq.~(\ref{SBL})
and
the three standard massive neutrinos are much lighter than the eV scale\footnote{
We do not consider the 1+3 scheme in which $\Delta m^2_{\text{SBL}}$
is obtained with a very light (or massless)
non-standard neutrinos and the three standard massive neutrinos have almost degenerate masses at the eV scale,
because this possibility is strongly disfavored by cosmological measurements
\cite{Ade:2015xua}
and by the experimental bounds on
neutrinoless double-$\beta$ decay
if the massive neutrinos are Majorana particles
(see Refs.~\cite{Bilenky:2014uka,Dell'Oro:2016dbc}).
}.
Let us emphasize that the 3+1 mixing scheme must be considered as effective,
in the sense that the existence of more non-standard massive neutrinos
is allowed, as long as their mixing with the three active neutrinos is sufficiently small
to be negligible in the analysis of the data of current experiments.

In the case of 3+1 neutrino mixing
\cite{Okada:1996kw,Bilenky:1996rw,Bilenky:1999ny,Maltoni:2004ei},
we have
$\Delta{m}^2_{41} = \Delta{m}^2_{\text{SBL}}$.
The oscillation probabilities of the flavor neutrinos in short-baseline experiments
are given by
\begin{equation}
P_{\nua{\alpha}\to\nua{\beta}}^{(\text{SBL})}
\simeq
\sin^2 2\vartheta_{\alpha\beta}
\sin^{2}\!\left( \frac{\Delta{m}^2_{41}L}{4E} \right)
\quad
(\alpha\neq\beta)
,
\qquad
P_{\nua{\alpha}\to\nua{\alpha}}^{(\text{SBL})}
\simeq
1
-
\sin^2 2\vartheta_{\alpha\alpha}
\sin^{2}\!\left( \frac{\Delta{m}^2_{41}L}{4E} \right)
,
\label{pro3p1}
\end{equation}
where $L$ is the source-detector distance and $E$ is the neutrino energy.
The oscillation amplitudes depend only
on the absolute values of the elements in the fourth column of the mixing matrix:
\begin{equation}
\sin^2 2\vartheta_{\alpha\beta}
=
4 |U_{\alpha 4}|^2 |U_{\beta 4}|^2
\quad
(\alpha\neq\beta)
,
\qquad
\sin^2 2\vartheta_{\alpha\alpha}
=
4
|U_{\alpha 4}|^2
\left(1 -  |U_{\alpha 4}|^2 \right)
.
\label{amp3p1}
\end{equation}
Hence, even if there are CP-violating phases in the mixing matrix,
CP violation cannot be measured in short-baseline experiments.
However,
the effects of the non-standard CP-violating phases
are observable in the experiments sensitive to the oscillations generated by the smaller squared-mass differences
$\Delta{m}^2_{\text{ATM}}$
\cite{deGouvea:2014aoa,Klop:2014ima,Berryman:2015nua,Gandhi:2015xza,Palazzo:2015gja,Agarwalla:2016mrc,Agarwalla:2016xxa,Dutta:2016glq}
and
$\Delta{m}^2_{\text{SOL}}$
\cite{Long:2013hwa}.
The dependence on the same elements of the mixing matrix
of the different amplitudes of the oscillations in short-baseline
appearance and disappearance experiments
generates the appearance-disappearance constraint
\cite{Okada:1996kw,Bilenky:1996rw}
\begin{equation}
\sin^2 2\vartheta_{\alpha\beta}
\simeq
\frac{1}{4}
\,
\sin^2 2\vartheta_{\alpha\alpha}
\,
\sin^2 2\vartheta_{\beta\beta}
\qquad
(\alpha=e,\mu,\tau)
.
\label{appdis3p1}
\end{equation}

The most recent global fits of short-baseline neutrino oscillation data
\cite{Gariazzo:2015rra,Collin:2016aqd,Collin:2016rao}
indicate that the most likely values of the 3+1 mixing parameters lie
in a region around
\begin{equation}
\Delta{m}^2_{41} \approx 1 \text{-} 2 \, \text{eV}^2
,
\quad
|U_{e4}|^2 \approx 0.03
,
\quad
|U_{\mu4}|^2 \approx 0.01
.
\label{3p1bf}
\end{equation}

\begin{figure}[t!]
\begin{center}
\begin{tabular}{cc}
\subfigure[]{\label{fig:gloa}
\includegraphics*[width=0.49\linewidth]{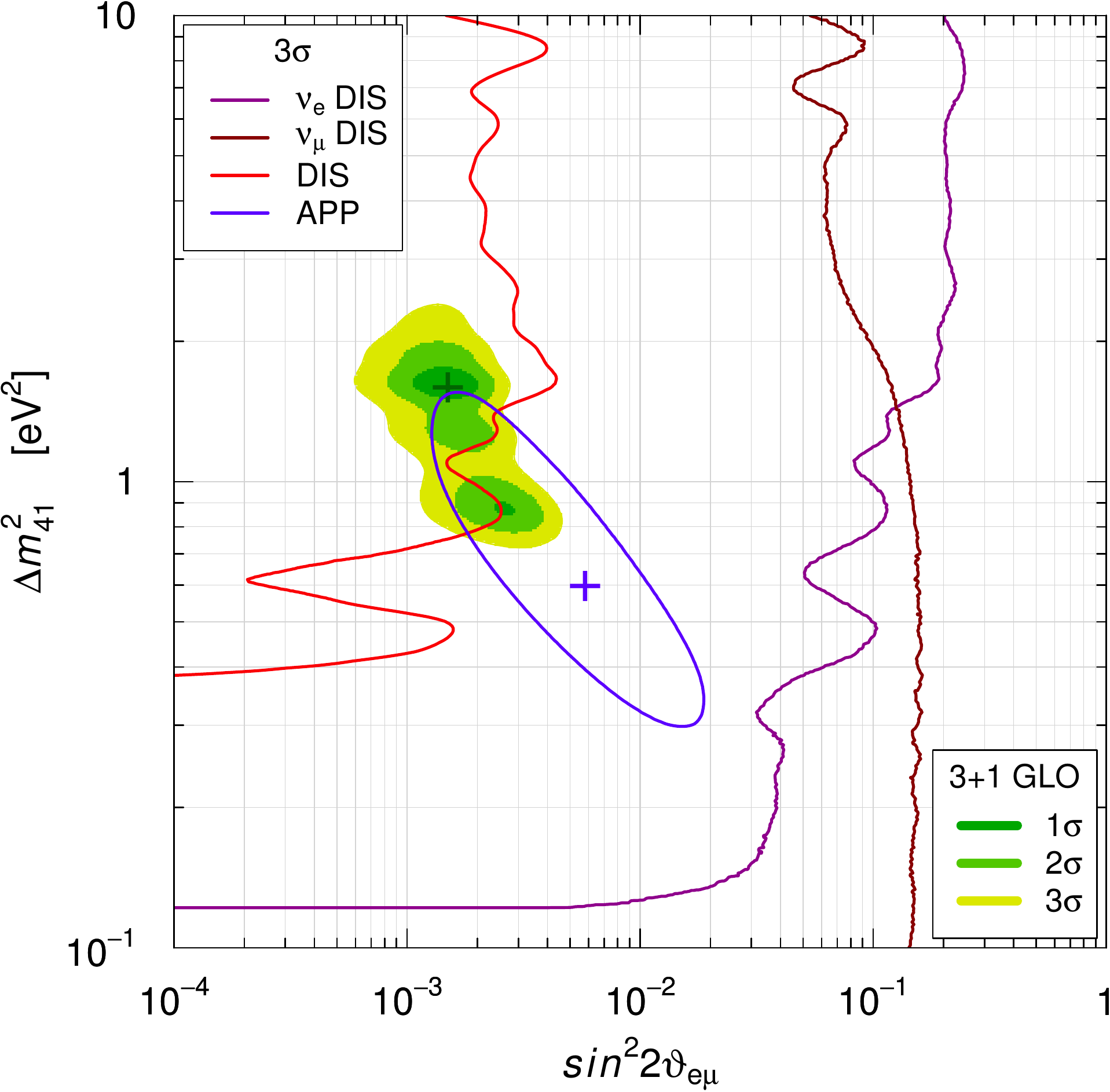}
}
&
\subfigure[]{\label{fig:glob}
\includegraphics*[width=0.49\linewidth]{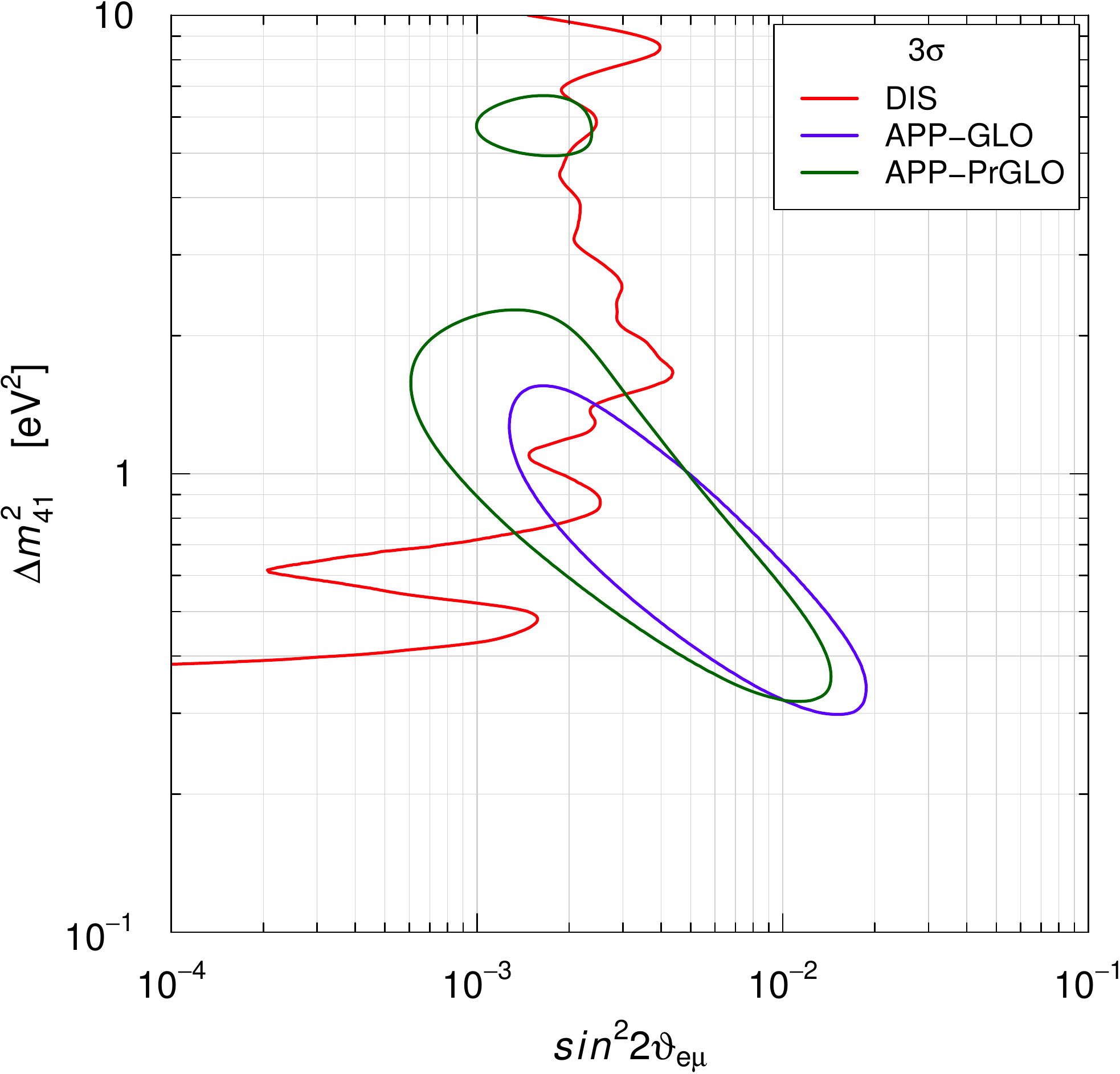}
}
\end{tabular}
\end{center}
\vspace{-0.7cm}
\caption{ \label{fig:glo}
\subref{fig:gloa}
Allowed regions in the
$\sin^{2}2\vartheta_{e\mu}$--$\Delta{m}^{2}_{41}$
plane
obtained in the 3+1 global (GLO) fit
of short-baseline neutrino oscillation data
compared with the $3\sigma$ allowed regions
obtained from
$\protect\nua{\mu}\to\protect\nua{e}$
short-baseline appearance data (APP)
and the $3\sigma$ constraints obtained from
$\protect\nua{e}$
short-baseline disappearance data ($\nu_{e}$ DIS),
$\protect\nua{\mu}$
short-baseline disappearance data ($\nu_{\mu}$ DIS)
and the
combined short-baseline disappearance data (DIS).
The best-fit points of the GLO and APP fits are indicated by crosses.
\subref{fig:glob}
Comparison of the
allowed regions obtained in the global (APP-GLO) and pragmatic (APP-PrGLO) fits
of short-baseline appearance data.
}
\end{figure}

\begin{figure}[t!]
\begin{center}
\begin{tabular}{cc}
\subfigure[]{\label{fig:CACS-160200671-3p1}
\includegraphics*[width=0.49\linewidth]{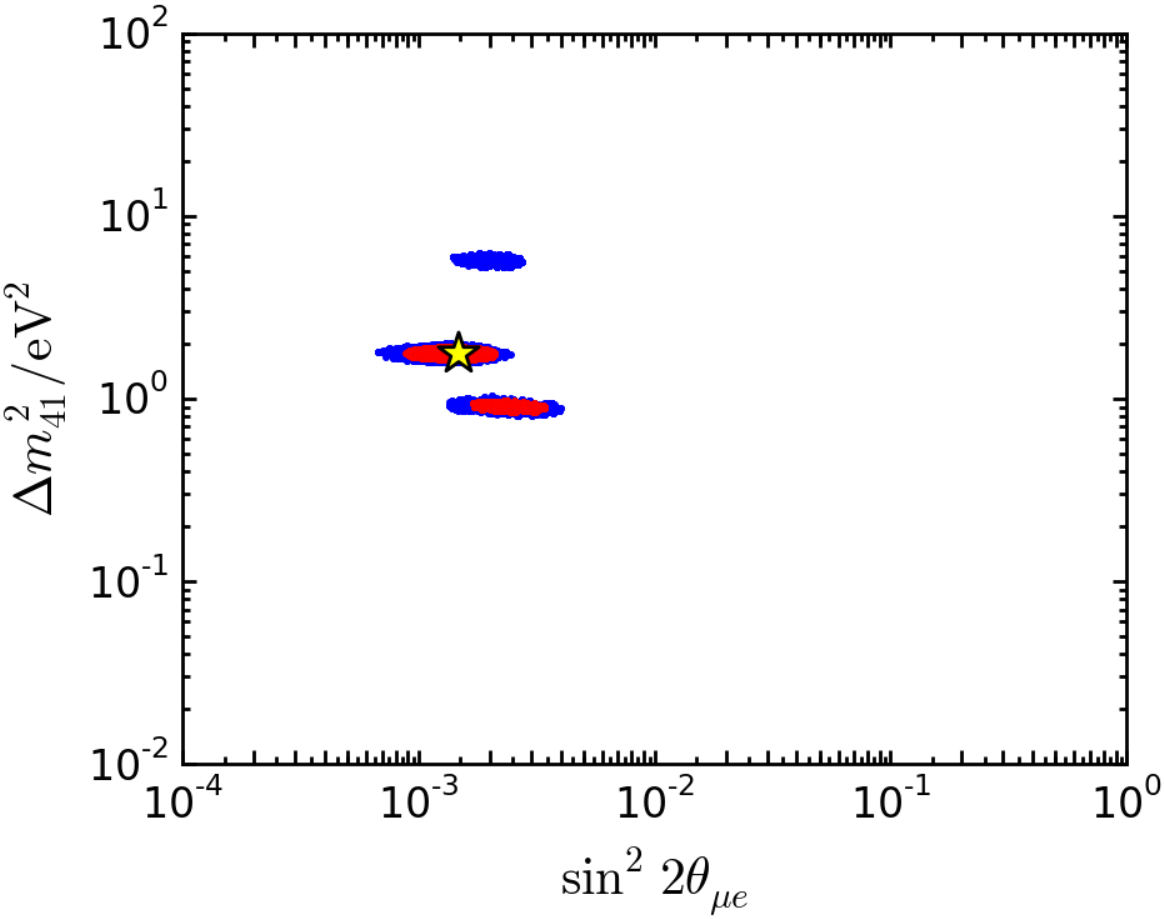}
}
&
\subfigure[]{\label{fig:CACS-IceCube-p3}
\includegraphics*[, viewport=60 573 285 749, width=0.49\linewidth]{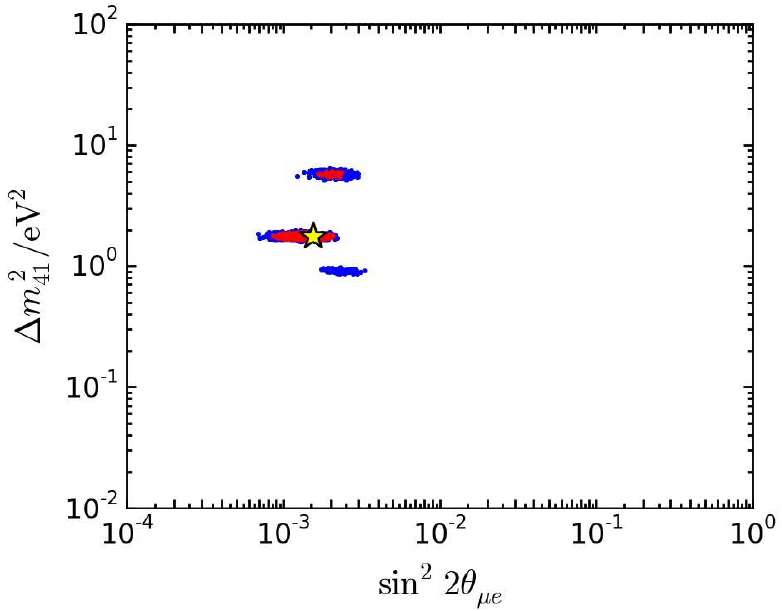}
}
\end{tabular}
\end{center}
\vspace{-0.7cm}
\caption{ \label{fig:CACS}
\subref{fig:gloa}
Allowed regions in the
$\sin^{2}2\vartheta_{e\mu}$--$\Delta{m}^{2}_{41}$
plane
obtained in the 3+1 global fits
of
Ref.~\cite{Collin:2016rao} \subref{fig:CACS-160200671-3p1}
and
Ref.~\cite{Collin:2016aqd} \subref{fig:CACS-IceCube-p3}.
The red and blue regions correspond, respectively to 90\% and 99\% C.L.
The best-fit point is marked by a yellow star.
}
\end{figure}

Figure~\ref{fig:gloa}
shows the allowed regions in the
$\sin^{2}2\vartheta_{e\mu}$--$\Delta{m}^{2}_{41}$
plane
obtained in the 3+1 global (GLO) fit
of short-baseline neutrino oscillation data
compared with the bounds from the data of disappearance experiments.
The best-fit values of the oscillation parameters are
$(\Delta{m}^{2}_{41})_{\text{bf}} = 1.6 \, \text{eV}^2$,
$(|U_{e4}|^2)_{\text{bf}} = 0.028$,
$(|U_{\mu4}|^2)_{\text{bf}} = 0.014$,
which imply
$(\sin^{2}2\vartheta_{e\mu})_{\text{bf}} = 0.0015$,
$(\sin^{2}2\vartheta_{ee})_{\text{bf}} = 0.11$ and
$(\sin^{2}2\vartheta_{\mu\mu})_{\text{bf}} = 0.054$.

Figure~\ref{fig:CACS-160200671-3p1} shows
the allowed regions in the
$\sin^{2}2\vartheta_{e\mu}$--$\Delta{m}^{2}_{41}$
plane
obtained in the 3+1 global fits
of
Ref.~\cite{Collin:2016rao}.
Comparing figures \ref{fig:gloa} and \ref{fig:CACS-160200671-3p1}
one can see that there is an approximate agreement of the results of the
two different global fits.
The differences are mainly due to different ways of analyzing
old data on which there is limited information.
The best-fit values of the oscillation parameters obtained in Ref.~\cite{Collin:2016rao} are
in approximate agreement with those above:
$(\Delta{m}^{2}_{41})_{\text{bf}} = 1.75 \, \text{eV}^2$,
$(|U_{e4}|^2)_{\text{bf}} = 0.027$,
$(|U_{\mu4}|^2)_{\text{bf}} = 0.014$,
which imply
$(\sin^{2}2\vartheta_{e\mu})_{\text{bf}} = 0.0015$,
$(\sin^{2}2\vartheta_{ee})_{\text{bf}} = 0.11$ and
$(\sin^{2}2\vartheta_{\mu\mu})_{\text{bf}} = 0.54$.

From Figure~\ref{fig:gloa}
one can see that the separate $3\sigma$ exclusion curves
obtained from
$\protect\nua{e}$
and
$\protect\nua{\mu}$
short-baseline disappearance data
do not exclude any area of the
region that is allowed at $3\sigma$ by the analysis of the
$\protect\nua{\mu}\to\protect\nua{e}$
short-baseline appearance data.
These bounds are simply obtained taking into account that
from the unitarity of the mixing matrix
$|U_{\alpha4}|^2 \leq 1 - |U_{\beta4}|^2$
for $\alpha\neq\beta$,
which implies that
$\sin^{2}2\vartheta_{\alpha\beta}
\leq
4 |U_{\alpha4}|^2 \left( 1 - |U_{\alpha4}|^2 \right)
=
\sin^{2}2\vartheta_{\alpha\alpha}$.
On the other hand,
when the
$\protect\nua{e}$
and
$\protect\nua{\mu}$
bounds
are combined through the appearance-disappearance constraint in Eq.~(\ref{appdis3p1})
the disappearance exclusion curve excludes most of the appearance allowed region.
Hence, there is a strong
appearance-disappearance tension\footnote{
This tension is unavoidable in any 3+$N_{s}$ scheme with $N_{s}$ sterile neutrinos
\cite{Giunti:2015mwa},
because the mixing of $\nu_{e}$ and $\nu_{\mu}$
with the sterile neutrinos required by the appearance data
implies $\nu_{e}$ and $\nu_{\mu}$ disappearances that are larger than
the respective experimental bounds.
}
\cite{Kopp:2011qd,Giunti:2011gz,Giunti:2011hn,Giunti:2011cp,Conrad:2012qt,Archidiacono:2012ri,Archidiacono:2013xxa,Kopp:2013vaa,Giunti:2013aea,Gariazzo:2015rra,Giunti:2015mwa}.

The appearance-disappearance tension
can be alleviated by excluding from the fit
the low-energy bins of the MiniBooNE experiment
\cite{AguilarArevalo:2008rc,Aguilar-Arevalo:2013pmq}
which have an anomalous excess of $\nu_{e}$-like events.
This is the pragmatic approach (PrGLO)
advocated in Ref.~\cite{Giunti:2013aea}.
The motivation is that
the MiniBooNE low-energy excess requires a small value of $\Delta{m}^2_{41}$
and a large value of $\sin^22\vartheta_{e\mu}$
\cite{Giunti:2011hn,Giunti:2011cp},
which are excluded by the data of other experiments
(see Ref.~\cite{Giunti:2013aea} for further details).
This is illustrated in Figure~\ref{fig:glob}
where one can see that the region allowed by appearance data
shifts towards larger values of $\Delta{m}^2_{41}$
and smaller values of $\sin^22\vartheta_{e\mu}$
when the MiniBooNE low-energy bins are omitted from the fit.
As a result, the overlap of the appearance and disappearance allowed regions
increases,
relieving the appearance-disappearance tension.
Therefore,
it is reasonable to adopt the pragmatic approach,
waiting for a clarification of the cause of the
MiniBooNE low-energy excess
by the MicroBooNE\footnote{
In the MiniBooNE  mineral-oil Cherenkov detector
$\nu_{e}$-induced events
cannot be distinguished from
$\nu_{\mu}$-induced events which produce only a visible photon
(for example neutral-current $\pi^{0}$ production in which only one of the two decay photons is visible).
On the other hand,
MicroBooNE is a large Liquid Argon Time Projection Chamber (LArTPC)
in which electrons and photons can be distinguished.
}
experiment at Fermilab
\cite{Gollapinni:2015lca}.

\begin{figure}[t!]
\begin{center}
\begin{tabular}{ccc}
\subfigure[]{\label{fig:prgloa}
\includegraphics*[width=0.3\linewidth]{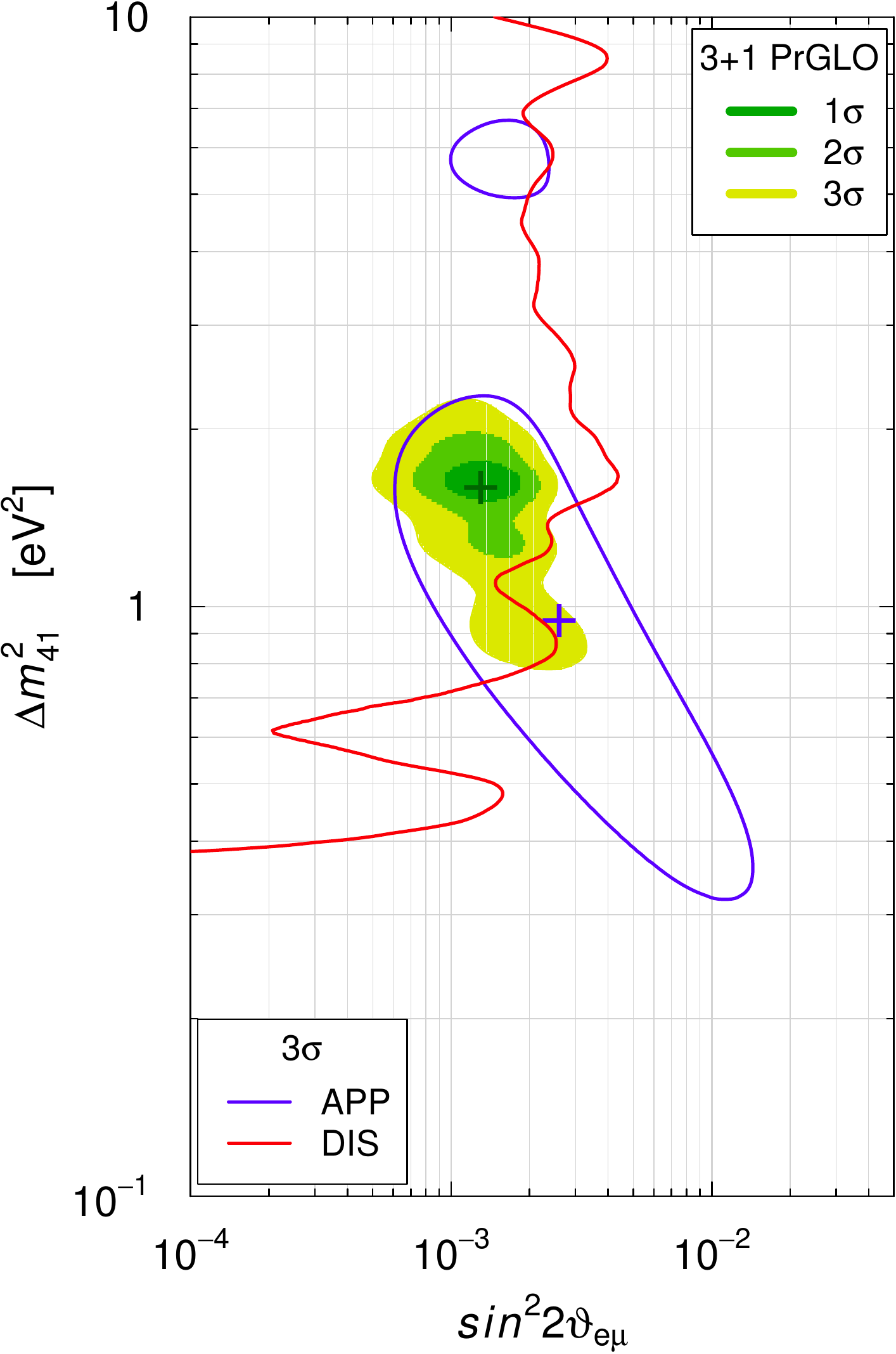}
}
&
\subfigure[]{\label{fig:prglob}
\includegraphics*[width=0.3\linewidth]{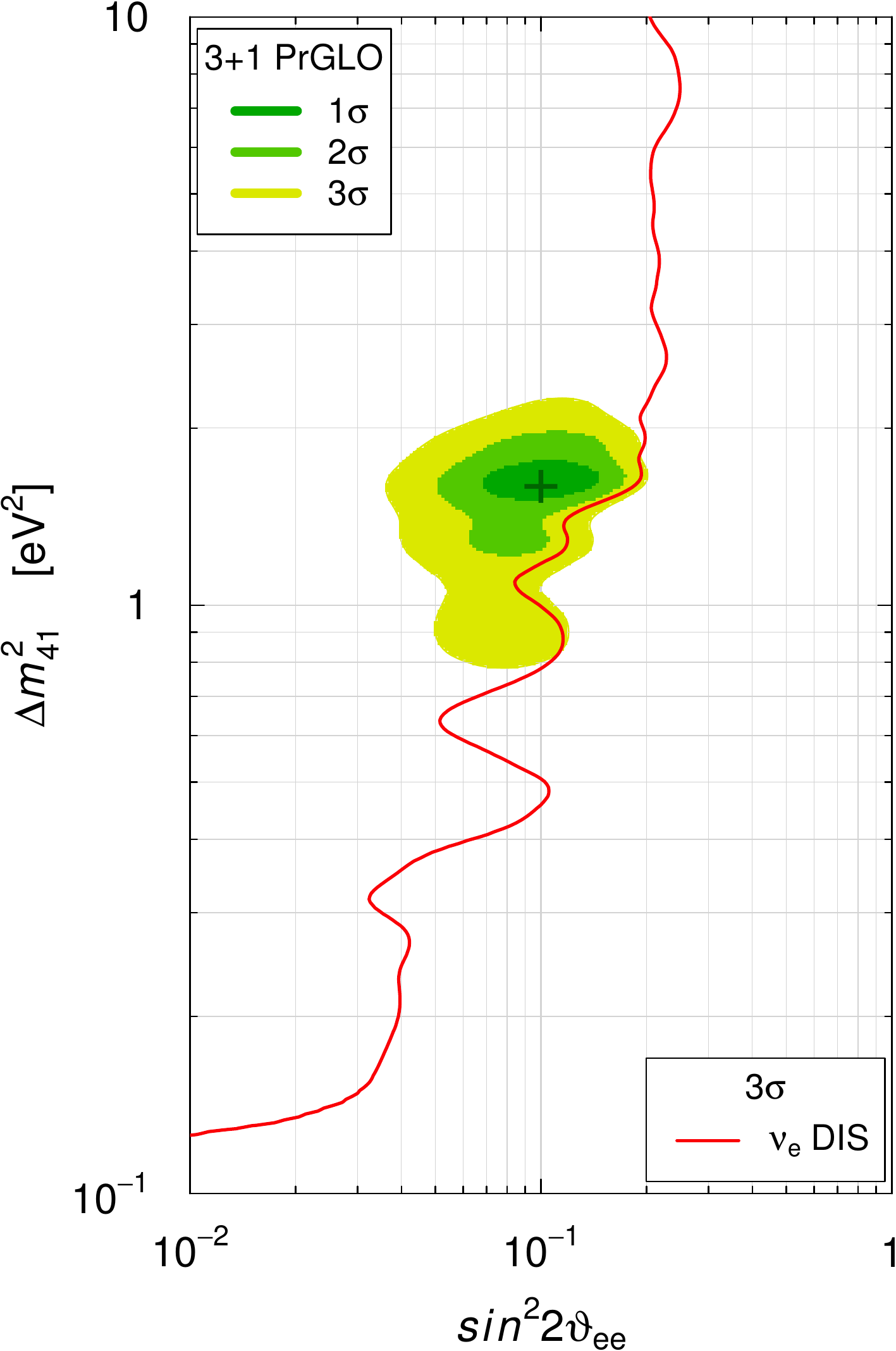}
}
&
\subfigure[]{\label{fig:prgloc}
\includegraphics*[width=0.3\linewidth]{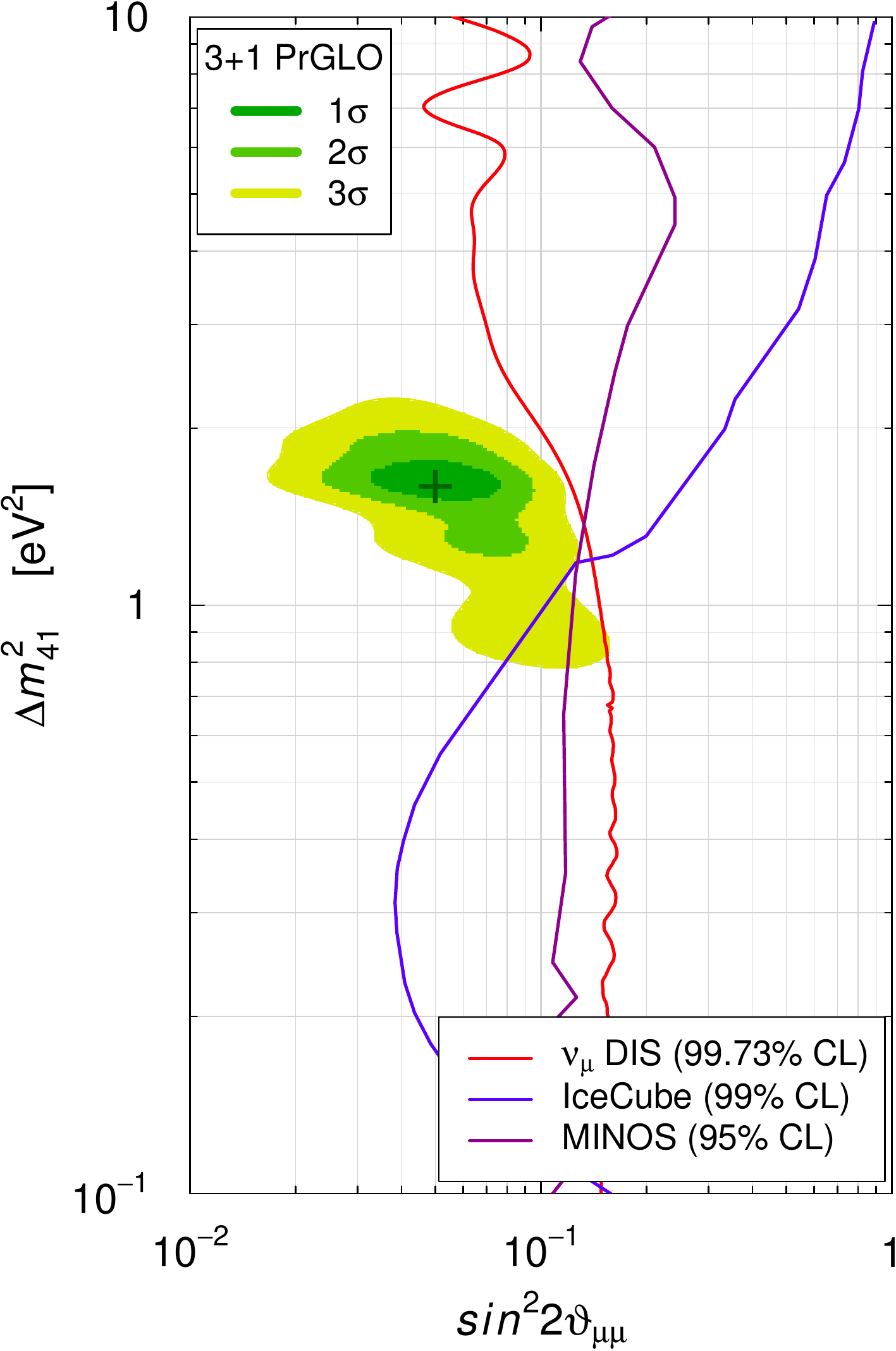}
}
\end{tabular}
\end{center}
\vspace{-0.7cm}
\caption{ \label{fig:prglo}
Allowed regions in the
$\sin^{2}2\vartheta_{e\mu}$--$\Delta{m}^{2}_{41}$ \subref{fig:prgloa},
$\sin^{2}2\vartheta_{ee}$--$\Delta{m}^{2}_{41}$ \subref{fig:prglob},
and
$\sin^{2}2\vartheta_{\mu\mu}$--$\Delta{m}^{2}_{41}$ \subref{fig:prgloc}
planes
obtained in the pragmatic 3+1 global fit PrGLO
of short-baseline neutrino oscillation data
compared with the $3\sigma$ allowed regions
obtained from
$\protect\nua{\mu}\to\protect\nua{e}$
short-baseline appearance data (APP)
and the $3\sigma$ constraints obtained from
$\protect\nua{e}$
short-baseline disappearance data ($\nu_{e}$ DIS),
$\protect\nua{\mu}$
short-baseline disappearance data ($\nu_{\mu}$ DIS)
and the
combined short-baseline disappearance data (DIS).
The best-fit points of the global (PrGLO) and APP fits are indicated by crosses.
}
\end{figure}

Figure~\ref{fig:prglo}
shows the allowed regions in the
$\sin^{2}2\vartheta_{e\mu}$--$\Delta{m}^{2}_{41}$,
$\sin^{2}2\vartheta_{ee}$--$\Delta{m}^{2}_{41}$ and
$\sin^{2}2\vartheta_{\mu\mu}$--$\Delta{m}^{2}_{41}$
planes
obtained from an update of the analysis in Ref.~\cite{Gariazzo:2015rra}
with an improved treatment of the MiniBooNE background disappearance
due to neutrino oscillations
\cite{Louis-private-16}.
These regions are relevant, respectively, for
$\nua{\mu}\to\nua{e}$ appearance,
$\nua{e}$ disappearance and
$\nua{\mu}$ disappearance
searches.
Figure~\ref{fig:prglo}
shows also the region allowed by $\nua{\mu}\to\nua{e}$ appearance data
and
the constraints from
$\nua{e}$ disappearance and
$\nua{\mu}$ disappearance data.
The best-fit values of the oscillation parameters are
$(\Delta{m}^{2}_{41})_{\text{bf}} = 1.6 \, \text{eV}^2$,
$(|U_{e4}|^2)_{\text{bf}} = 0.027$,
$(|U_{\mu4}|^2)_{\text{bf}} = 0.012$,
which imply
$(\sin^{2}2\vartheta_{e\mu})_{\text{bf}} = 0.0013$,
$(\sin^{2}2\vartheta_{ee})_{\text{bf}} = 0.10$ and
$(\sin^{2}2\vartheta_{\mu\mu})_{\text{bf}} = 0.050$.

Figure~\ref{fig:prgloc}
shows a comparison of the
allowed regions in the
$\sin^{2}2\vartheta_{\mu\mu}$--$\Delta{m}^{2}_{41}$
plane
with the exclusion curves
obtained recently by the
IceCube \cite{TheIceCube:2016oqi}
and
MINOS \cite{MINOS:2016viw}
experiments.
One can see that they disfavor
the low-$\Delta{m}^{2}_{41}$ and high-$\sin^{2}2\vartheta_{\mu\mu}$
part of the allowed region.
This is confirmed by the results presented in Ref.~\cite{Collin:2016aqd},
where the 3+1 global fit of
Ref.~\cite{Collin:2016rao}
was updated with the addition of the IceCube data.
The resulting allowed regions in the
$\sin^{2}2\vartheta_{\mu\mu}$--$\Delta{m}^{2}_{41}$
plane
are shown in Figure~\ref{fig:CACS-IceCube-p3}.
Comparing Figures \ref{fig:CACS-160200671-3p1} and \ref{fig:CACS-IceCube-p3}
one can see that the effect of including the IceCube data in the fit
is to disfavor
the low-$\Delta{m}^{2}_{41}$ region.
The main allowed region around the best-fit point remains stable
and
there is a slight improvement of the likelihood of the high-$\Delta{m}^{2}_{41}$ region.

\begin{figure}[t!]
\begin{center}
\begin{tabular}{ccc}
\subfigure[]{\label{fig:futurea}
\includegraphics*[width=0.3\linewidth]{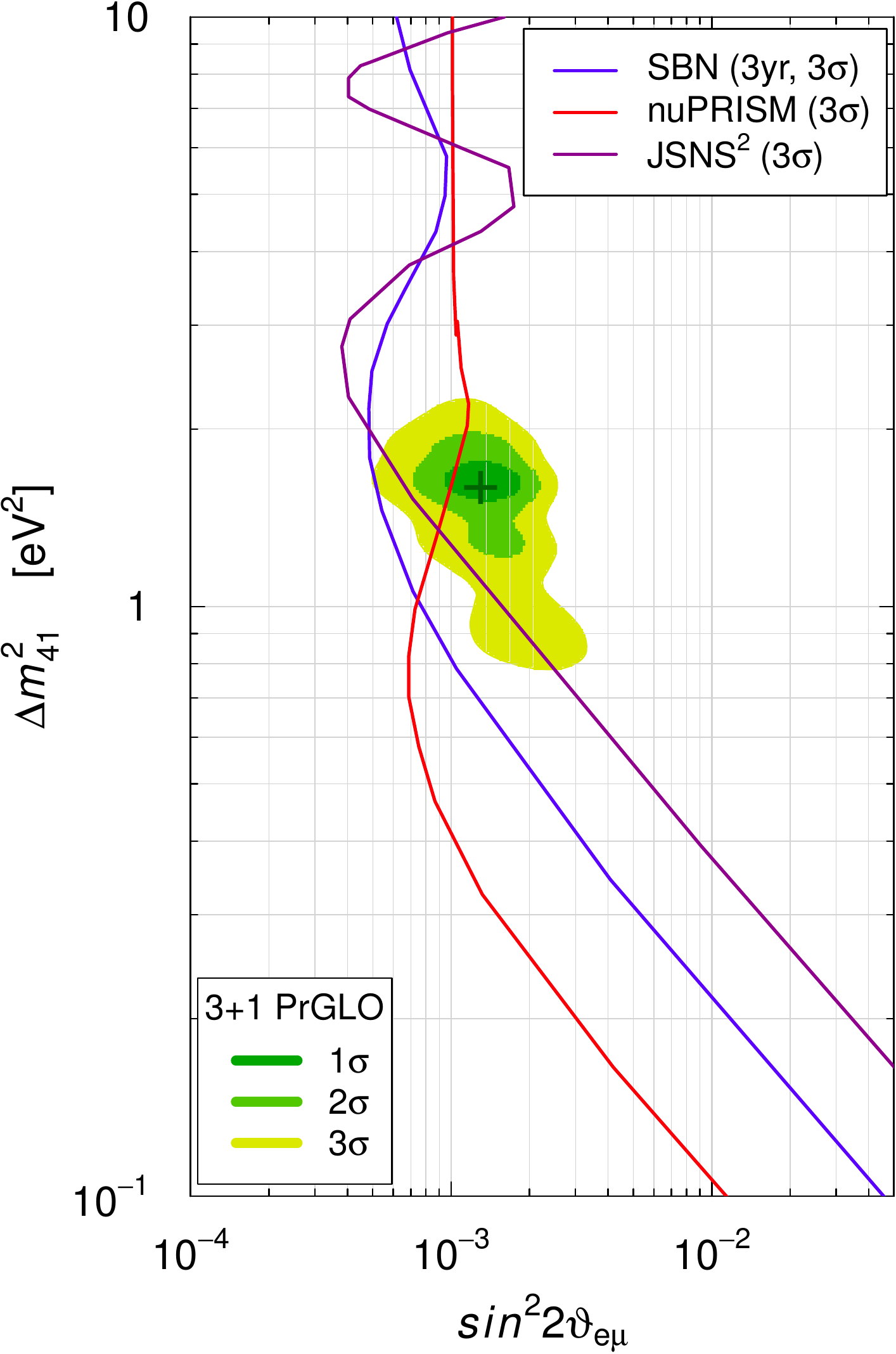}
}
&
\subfigure[]{\label{fig:futureb}
\includegraphics*[width=0.3\linewidth]{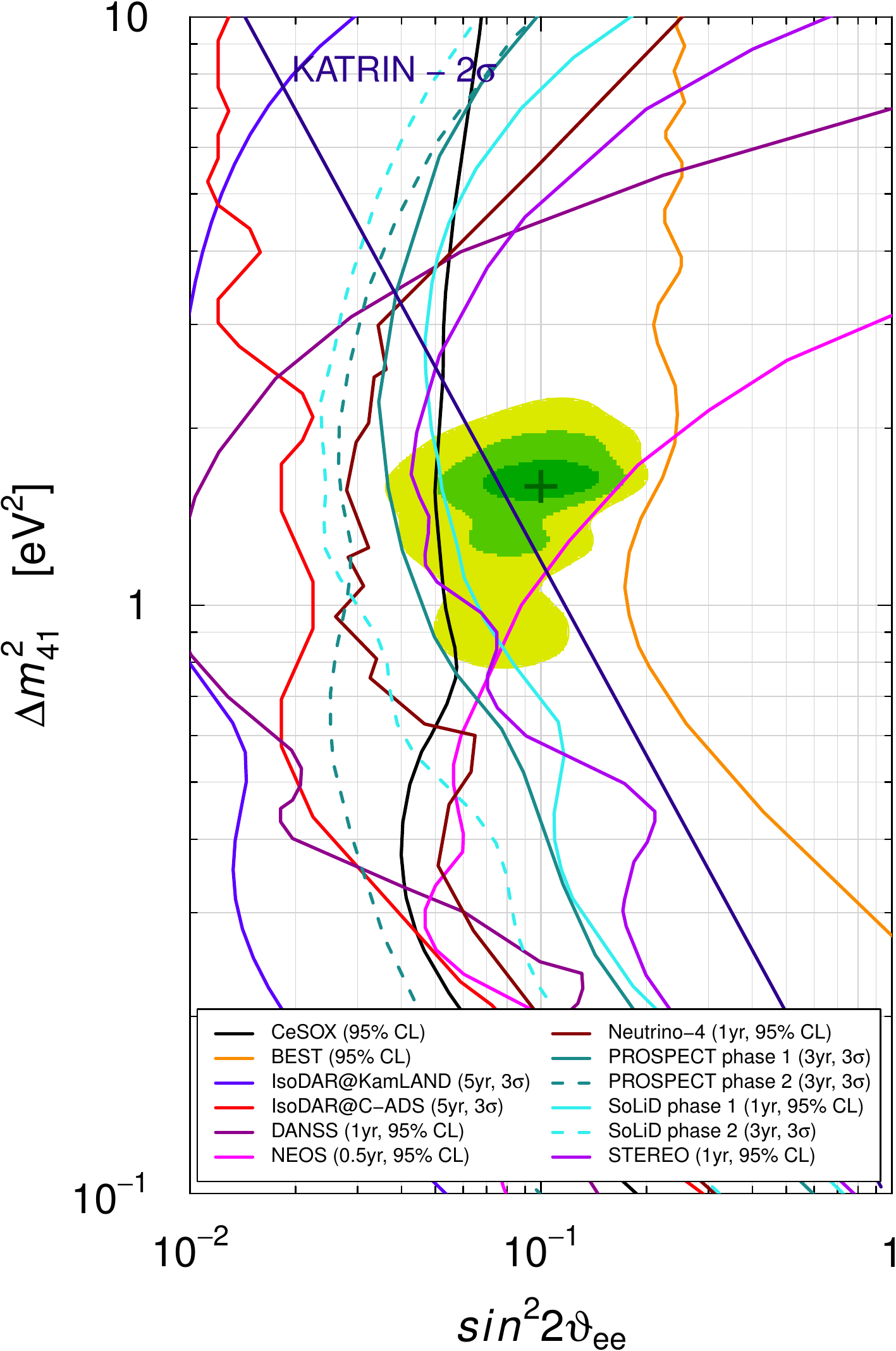}
}
&
\subfigure[]{\label{fig:futurec}
\includegraphics*[width=0.3\linewidth]{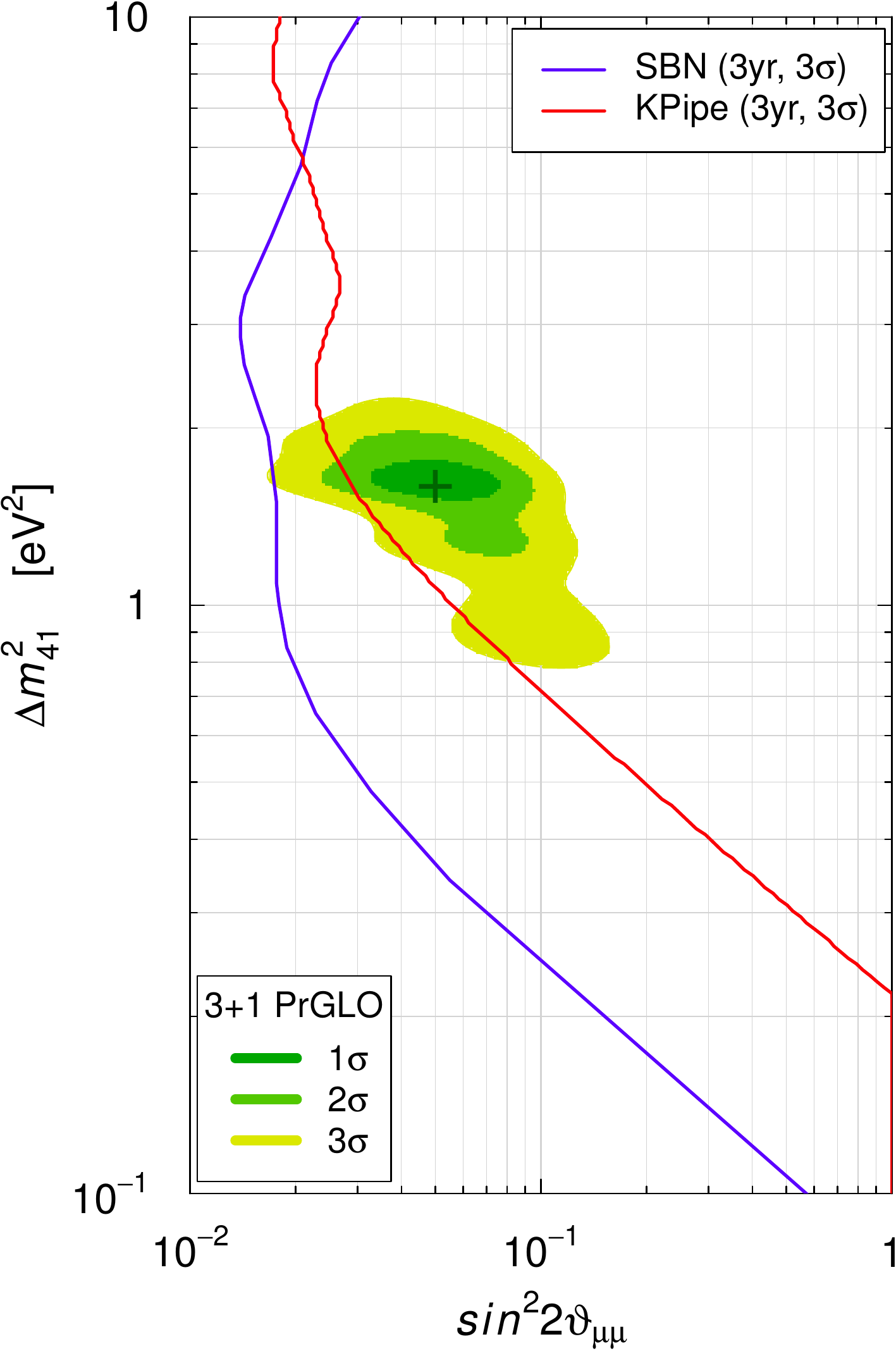}
}
\end{tabular}
\end{center}
\vspace{-0.7cm}
\caption{ \label{fig:future}
Sensitivities of future experiments
compared with the PrGLO allowed regions of Fig.~\ref{fig:prglo}.
}
\end{figure}

Because of the scarcity of sensitive data,
of the possible existence of unknown systematic errors,
and
of the appearance-disappearance tension,
the possible existence of light sterile neutrinos
at the eV scale
is controversial and needs new reliable experimental checks.
Fortunately,
there is an impressive program of new experiments
which are planned to check the existence
of eV sterile neutrinos
(see the reviews in Refs.~\cite{Gariazzo:2015rra,Giunti:2015wnd,Stanco:2016gnl,Fava:2016vas}).
Figure~\ref{fig:future}
shows a comparison of the sensitivities of future experiments
with the PrGLO allowed regions of Fig.~\ref{fig:prglo}
for
\subref{fig:futurea}
$\nua{\mu}\to\nua{e}$
transitions
(SBN \cite{Antonello:2015lea},
nuPRISM \cite{Bhadra:2014oma},
JSNS$^2$ \cite{Harada:2013yaa}),
\subref{fig:futureb}
$\nua{e}$
disappearance
(CeSOX \cite{Borexino:2013xxa},
BEST \cite{Barinov:2016znv},
IsoDAR@KamLAND \cite{Abs:2015tbh},
IsoDAR@C-ADS \cite{Ciuffoli:2015uta},
DANSS \cite{Alekseev:2016llm},
NEOS \cite{Kim:2015qlu},
Neutrino-4 \cite{Serebrov:2012sq},
PROSPECT \cite{Ashenfelter:2015uxt},
SoLid \cite{Ryder:2015sma},
STEREO \cite{Helaine:2016bmc},
KATRIN \cite{Mertens-TAUP2015}),
and
\subref{fig:futurec}
$\nua{\mu}$
disappearance
(SBN \cite{Antonello:2015lea},
KPipe \cite{Axani:2015zxa}).

Moreover,
light sterile neutrinos
have important effects that could be observed
in $\beta$ decay experiments
\cite{Riis:2010zm,Formaggio:2011jg,SejersenRiis:2011sj,Esmaili:2012vg,Gastaldo:2016kak},
in neutrinoless double-$\beta$ decay experiments
\cite{Barry:2011wb,Li:2011ss,Rodejohann:2012xd,Giunti:2012tn,Girardi:2013zra,Pascoli:2013fiz,Meroni:2014tba,Abada:2014nwa,Giunti:2015kza,Pas:2015eia},
in solar neutrino experiments
\cite{Palazzo:2011rj,Palazzo:2012yf,Giunti:2012tn,Palazzo:2013me,Long:2013hwa,Long:2013ota,Kopp:2013vaa},
in long-baseline neutrino oscillation experiments
\cite{deGouvea:2014aoa,Klop:2014ima,Berryman:2015nua,Gandhi:2015xza,Palazzo:2015gja,Agarwalla:2016mrc,Agarwalla:2016xxa,Choubey:2016fpi,Agarwalla:2016xlg},
in atmospheric neutrino experiments
\cite{Razzaque:2011ab,Razzaque:2012tp,Gandhi:2011jg,Esmaili:2012nz,Esmaili:2013cja,Esmaili:2013vza,Rajpoot:2013dha,Lindner:2015iaa,Liao:2016reh},
in supernova neutrino experiments
\cite{Tamborra:2011is,Wu:2013gxa,Esmaili:2014gya,Warren:2014qza},
in indirect dark matter detection
\cite{Esmaili:2012ut}),
in high-energy cosmic neutrinos experiments
\cite{Barry:2010en},
and in cosmology
(see Refs.~\cite{Archidiacono:2013fha,Lesgourgues:2014zoa,Gariazzo:2015rra}).

Let us finally emphasize that the discovery of the existence of sterile neutrinos would be a major discovery
which would have a profound impact not only on neutrino physics,
but on our whole view of fundamental physics,
because sterile neutrinos are elementary particles beyond the Standard Model.
The existence of light sterile neutrinos would prove that there is new physics beyond the Standard Model at low-energies
and their properties can give important information on this new physics
(see Refs.~\cite{Volkas:2001zb,Mohapatra:2006gs}).


\end{document}